\documentstyle[11pt,amssymb,epsfig,psfig]{article}

\begin{document}

\title{ Surface vacuum energy and stresses for a brane in de Sitter spacetime }

\author{
 M. R. Setare  \footnote{E-mail: rezakord@ipm.ir} \\
  { Physics Department, Institute for Studies in Theoretical Physics and }
\\ {Mathematics, Tehran, Iran} }

\maketitle

\begin{abstract}
Vacuum expectation values of the surface energy-momentum tensor
is investigated for a massless scalar field obeying mixed
boundary condition on a brane in de Sitter bulk. To generate the
corresponding vacuum surface densities we use the conformal
relation between de Sitter and Rindler spacetimes.
\end{abstract}

\bigskip

{PACS number(s): 03.70.+k, 11.10.Kk}

\newpage

\section{Introduction} \label{sec:int}

De Sitter (dS) spacetime is the maximally symmetric solution of Einsten's equation
with a positive cosmological constant. Recent astronomical observations of supernovae
and cosmic microwave background \cite{Ries98} indicate that the universe is accelerating
 and can be well approximated by a world with a positive cosmological constant. If the
  universe would accelerate indefinitely, the standard cosmology leads to an asymptotic dS
   universe. De Sitter spacetime plays an important role in the inflationary scenario,
   where an exponentially expanding approximately dS spacetime is employed to solve
    a number of problems in standard cosmology. The quantum field theory on dS spacetime
    is also of considerable interest. In particular, the inhomogeneities generated by
    fluctuations of a quantum field during inflation provide an attractive mechanism for
     the structure formation in the universe. Another motivation for investigations of
     dS based quantum theories is related to the recently proposed holographic duality between
     quantum gravity on dS spacetime and a quantum field theory living on boundary identified
     with the timelike infinity of dS spacetime \cite{Stro01}.\\
     The investigation of quantum effects in braneworld models is
     of considerable phonomenological interest, both in particle
     physics and in cosmology. The braneworld corresponds to a
     manifold with dynamical boundaries and all fields which
     propagate in the bulk will give Casimir-type contributions to the vacuum energy, and
     as a result to the vacuum forces acting on the branes. In
     dependence of the type of a field and boundary conditions
     imposed, these forces can either stabilize or destabilize the
     braneworld. In addition, the Casimir energy gives a
     contribution to both the brane and bulk cosmological
     constants and, hence, has to be taken into account in the self-consistent formulation of the
     braneworld dynamics. Motivated by these, the role of quantum
     effects in braneworld scenarios has received a grate deal of
     attention. For conformally coupled scalar this effect was
     initially studied in \cite{fab} in the context of M-theory,
     and subsequently in \cite{odi}-\cite{nor} for a background
     Randall-Sundrum geometry. The models with dS and AdS brane,
     and higher dimensional brane models are considered as well
     \cite{Eliz03}, \cite{Noji00}-\cite{nor}. For a conformally
     coupled bulk scalar the cosmological backreaction of the
     Casimir energy is investigated in \cite{fab},\cite{Eliz03},
     \cite{Noji00}.

 As a brane we take $4$-dimensional hypersurface
   which is the conformal image of a plate moving with constant proper acceleration in the
    Rindler stacetime. We will assume that the field is prepared in the state
    conformally related to the Fulling--Rindler vacuum in the Rindler spacetime.
    To generate the vacuum expectation values in dS bulk, we use the conformal
    relation between dS and Rindler spacetimes and the results from \cite{setsahr}
   for the corresponding Rindler problem with mixed boundary conditions. Previously
   this method has been used in \cite{set5} to derive the vacuum stress on parallel plates
   for a scalar field with Dirichlet boundary conditions in de Sitter spactime and in Ref. \cite{set6}
to investigate the vacuum characteristics of the Casimir configuration
on background of conformally flat brane-world geometries for
massless scalar field with Robin boundary conditions on plates.

The present paper is organized as follows. In the next section the geometry of our problem and the conformal relation between dS and Rindler spacetimes are discussed. The results are presented for the vacuum expectation values of the energy-momentum tensor for a scalar field induced by a plate uniformly accelerated through the Fulling--Rindler vacuum. In Section \ref{sec2}, by using the formula relating the renormalized energy-momentum tensors for conformally related problems in combination with the appropriate coordinate transformation, we derive expressions for the vacuum energy-momentum tensor in dS space. The main results are rementioned and summarized in Section \ref{secconc}.

\section{Conformal relation between dS and Rindler problems} \label{sec1}

Consider a conformally coupled massless scalar field $%
\varphi (x)$ satisfying the equation
\begin{equation}
\left( \nabla _{l}\nabla ^{l}+\zeta R\right) \varphi (x)=0,\quad
\zeta =\frac{3}{16} , \label{fieldeq}
\end{equation}
on background of a $4+1$--dimensional dS spacetime. In Eq.
(\ref{fieldeq}), $\nabla _{l}$ is the operator of the covariant
derivative, and $R$ is the Ricci scalar for the corresponding
metric $g_{ik}$. In static coordinates $x^i=(t,r,\theta ,\theta
_2,\phi )$ dS metric has the form
\begin{equation}
ds_{{\rm dS}}^{2}=g_{ik}dx^idx^k=\left( 1-\frac{r^{2}}{\alpha
^{2}}\right) dt^{2}-\frac{dr^{2}}{1-\frac{r^{2}}{\alpha
^{2}}}-r^{2}d\Omega ^{2}_{3} , \label{ds2dS}
\end{equation}
where $d\Omega ^{2}_{3}$ is the line element on the
$3$--dimensional unit sphere in Euclidean space, and the
parameter $\alpha $ defines the dS curvature radius. Note that
$R=12/\alpha ^2$. We will assume that the field satisfies the
mixed boundary condition
\begin{equation}
\left( A_s+n^{l}\nabla _{l}\right) \varphi (x)=0,\quad  ,
\label{boundcond}
\end{equation}
on the brane, where $A_s$ is a constant and $n^{l}$ is the unit
inward normal to the brane. This type of conditions is an

extension of Dirichlet and Neumann boundary conditions and appears

in a variety of situations, for example the casimir effect for
massless scalar filed with Robin boundary conditions on two
parallel plates in de Sitter spactime is calculated in
\cite{set5}, the Robin type boundary condition in domain wall
formation is investigated in \cite{setd}.
 Mixed boundary conditions naturally

arise for scalar and fermion bulk fields in the Randall-Sundrum

model \cite{fla2,set6, sahari}.
 To
make maximum use of the flat spacetime calculations, first of all
let us present the dS line element in the form conformally
related to the Rindler metric. With this aim we make the
coordinate transformation $x^i\to x'^{i}=(\tau ,\xi
,{\mathbf{x}}' )$, ${\mathbf{x}}'=(x'^{2},x'^{3} ,x'^{4})$
\begin{eqnarray} \label{coord}
&& \tau =\frac{t}{\alpha },\quad \xi =\frac{\sqrt{\alpha ^2-r^2}
}{\Omega },\quad x'^{2}=\frac{r}{\Omega }\sin \theta \cos \theta
_2 ,\nonumber \\
&&    x'^{3}=\frac{r}{\Omega }\sin \theta \sin \theta _2\cos \phi,
\quad x'^{4}=\frac{r}{\Omega }\sin \theta \sin \theta _2 \sin
\theta _{2}\sin \phi,
\end{eqnarray}
with the notation
\begin{equation}\label{Omega}
  \Omega =1-\frac{r}{\alpha }\cos \theta .
\end{equation}
Under this coordinate transformation the dS line element takes the
form
\begin{equation}
ds_{{\rm dS}}^{2}=g'_{ik}dx'^idx'^k=\Omega ^2\left( \xi ^{2}d\tau
^{2}-d\xi ^{2}-d{\mathbf{x}}'^{2}\right) .  \label{ds2dS1}
\end{equation}
In this form the dS metric is manifestly conformally related to
the Rindler spacetime with the line element $ds_{{\rm R}}^{2}$:
\begin{equation}
ds_{{\rm dS}}^{2}=\Omega ^{2}ds_{{\rm R}}^{2},\quad ds_{{\rm
R}}^{2}=g^{{\mathrm{R}}}_{ik}dx'^idx'^k=\xi ^{2}d\tau ^{2}-d\xi
^{2}-d{\mathbf{x}}'^{2},\quad g'_{ik}=\Omega ^2
g^{{\mathrm{R}}}_{ik}. \label{confrel}
\end{equation}
By using the standard transformation formula for the vacuum
expectation values of the energy-momentum tensor in conformally
related problems (see, for instance, \cite{Birrell}), we can
generate the results for dS spacetime from the corresponding
results for the Rindler spacetime. In this letter as a Rindler
counterpart we will take the vacuum surface energy-momentum
tensor induced by an infinite plate moving by uniform proper
acceleration through the Fulling-Rindler vacuum. We will assume
that the plate is located in the right Rindler wedge and has the
coordinate $\xi=a$. Observe that in coordinates $x^{i}$ the
boundary $\xi=a$ is presented by the hypersurface
 \begin{equation}
 \sqrt{\alpha^{2}-r^{2}}=a(1-\frac{r}{\alpha}\cos\theta),
 \label{sur}
 \end{equation}
 in dS spacetime.
 In Ref. \cite{Saha03emt} it was
argued that the energy-momentum tensor for a scalar field on
manifolds with boundaries in addition to the bulk part contains a
contribution located on the boundary. For the boundary $\partial
M_{s}$ the surface part of the energy-momentum tensor is presented
in the form \cite{Saha03emt}
\begin{equation}
T_{ik}^{{\rm (surf)}}=\delta (x;\partial M_{s})\tau _{ik}
\label{Ttausurf}
\end{equation}%
with
\begin{equation}
\tau _{ik}=\zeta \varphi ^{2}K_{ik}-(2\zeta -1/2)h_{ik}\varphi
n^{l}\nabla _{l}\varphi ,  \label{tausurf}
\end{equation}%

and the "one-sided" delta-function $\delta (x;\partial M_{s})$
locates this tensor on $\partial M_{s}$. In Eq. (\ref{tausurf}),
$K_{ik}$ is the extrinsic curvature tensor of the boundary
$\partial M_{s}$ and $h_{ik}$ is the corresponding induced
metric. Let $\{\varphi _{\alpha }(x),\varphi _{\alpha }^{\ast
}(x)\}$ be a complete set of positive and negative frequency
solutions to the field equation (\ref{fieldeq}), obeying boundary
condition (\ref{boundcond}). Here $\alpha $ denotes a set of
quantum numbers specifying the solution. By expanding the field
operator over the eigenfunctions $\varphi _{\alpha }(x)$, using
the standard commutation rules and the definition of the vacuum
state, for the vacuum expectation value of the surface
energy-momentum tensor one finds
\begin{equation}
\langle 0|T_{ik}^{{\rm (surf)}}|0\rangle =\delta (x;\partial
M_{s})\langle 0|\tau _{ik}|0\rangle ,\quad \langle 0|\tau
_{ik}|0\rangle =\sum_{\alpha }\tau _{ik}\{\varphi _{\alpha
}(x),\varphi _{\alpha }^{\ast }(x)\}, \label{modesumform}
\end{equation}%
where $|0\rangle $ is the amplitude for the corresponding vacuum
state, and the bilinear form $\tau _{ik}\{\varphi ,\psi \}$ on
the right of the second
formula is determined by the classical energy-momentum tensor (\ref{tausurf}%
). The surface energy-momentum tensor has a diagonal structure:
\begin{equation}
\langle 0|\tau _{l}^{k}|0\rangle ={\rm diag}\left( \varepsilon
,0,-p,-p,-p\right) ,  \label{taudiag}
\end{equation}%
with the surface energy density $\varepsilon $ and stress $p$,
and the
equation of state%
\begin{equation}
\varepsilon =-\left[ 1+\frac{2\zeta }{A(4\zeta -1)}\right] p,
\hspace{1cm}A=aA_s \label{eqstate}
\end{equation}%
Here we take a plane boundary with coordinate $\xi=a>0$
corresponding to a plate uniformly accelerated normal to itself
with the proper acceleration $a^{-1}$. For a minimally coupled
scalar field, $\varepsilon$ corresponds to a cosmological
constant induced on the plate. In the conformally coupled case
\begin{equation}
\varepsilon=-(1-\frac{3}{2A})p. \label{concas}
\end{equation}
 The vacuum stress induced on the
brane is as following \cite{setsahr}

\begin{equation}
p=p_{p}^{{\rm (R)}}+p_{f}^{{\rm (R)}},  \label{ppf}
\end{equation}%
where for the pole and finite contributions one has%
\begin{eqnarray}
p_{p}^{{\rm (R)}} &=&\frac{B_{4}}{4sa^{4}}AF_{{\rm R}%
,-1}^{(as)},\quad A=aA_{s},  \label{ppf1} \\
p_{f}^{{\rm (R)}} &=&\frac{B_{4}}{4a^{4}}A\left[ F_{{\rm R}%
,0}^{(as)}+F_{{\rm R}}^{(1)}(0)\right] ,
\end{eqnarray}%
and the coefficients are defined by following expressions
\begin{equation}
B_4=\frac{1}{(4\pi)^{3/2}\Gamma(3/2) }, \label{beq}
\end{equation}
\begin{eqnarray}
F_{{\rm R}}^{(1)}(s) &=&-\frac{1}{\pi }\cos \frac{\pi
s}{2}\int_{0}^{\infty
}dx\,x^{2}\int_{\rho }^{\infty }dz\,z^{-s}\left[ \frac{K_{z}(x)}{\bar{K}%
_{z}(x)}+\frac{1}{r}\sum_{l=0}^{N}\frac{(-1)^{l}U_{l}(\cos \theta )}{%
(1+r^{2})^{l/2}}\right] ,  \label{F1}
\end{eqnarray}%
\begin{equation}
F_{{\rm R},-1}^{(as)}=-\frac{2}{\pi }\Gamma \left( \frac{3}{2}%
\right) \sum_{j=0}^{1}\frac{(-1)^{j}}{\Gamma (j+1)\Gamma \left( \frac{3%
}{2}-j\right) }\sum_{m=0}^{3-2j}U_{3-2j,m}B\left( m+\frac{1}{2},\frac{3%
}{2}\right) ,  \label{Fas-1}
\end{equation}%

\begin{eqnarray}
F_{{\rm R},0}^{(as)} &=&-\frac{1}{\pi }\Gamma \left( \frac{3}{2}%
\right) \sum_{j=0}^{1}\frac{(-1)^{j}}{\Gamma (j+1)\Gamma \left( \frac{3%
}{2}-j\right) }\sum_{m=0}^{3-2j}U_{3-2j,m}B\left( m+\frac{1}{2},\frac{3%
}{2}\right)   \nonumber \\
&\times &\left[ \psi \left( m+2\right) +\psi \left( j+1\right)
-\psi \left( m+\frac{1}{2}\right) -\psi \left( \frac{3}{2}\right)
\right]
\label{Fas0} \\
&+&\frac{1}{\pi }\left( \sum_{l=1,4-l={\rm even}}^{3}+\sum_{l=4}^{N}%
\right) (-1)^{l}B\left( \frac{l-3}{2},\frac{3}{2}\right)
\sum_{m=0}^{l}U_{lm}B\left( m+\frac{1}{2},\frac{3}{2}\right) ,
\nonumber
\end{eqnarray}%
where $\psi (x)=d\ln \Gamma (x)/dx$ is the diagamma function and $B(x,y)$ is the beta function.\\
The surface energy per unit surface of the plate can be found
integrating the energy
density from Eq. (\ref{modesumform}),%
\begin{equation}
E^{{\rm (R,surf)}}=\int d^{4}x\sqrt{|\det g_{ik}|}\langle 0|T_{0}^{{\rm %
(surf)}0}|0\rangle =a\langle 0|\tau _{0}^{0}|0\rangle
=a\varepsilon . \label{ERsurf}
\end{equation}%
\section{Vacuum energy-momentum tensor in dS bulk}
\label{sec2}

To find the VEV's induced by the surface (\ref{sur}) in dS
spacetime, first we will consider the corresponding quantities in
the coordinates $(\tau ,\xi ,{\mathbf{x}}')$ with metric
(\ref{ds2dS1}). These quantities can be found from the
corresponding results in the Rindler spacetime by using the
standard transformation formula for the conformally related
problems \cite{Birrell}:

\begin{eqnarray}
\left\langle T_{i}^{k}\left[ g'_{lm},\varphi \right]\right\rangle
 &=&\Omega ^{-5}\left\langle
T_{i}^{k}[g^{{\mathrm{R}}}_{lm},\varphi
_{{\mathrm{R}}}]\right\rangle . \label{TikdSb}
\end{eqnarray}
If the classical energy-momentum tensor is traceless then the
classical action is invariant under the conformal transformation.
It must be noted that trace anomalies in stress tensor i.e. the
nonvanishing $T^{i}_{i}$ for a conformaly invariant field after
renormalization originate from some quantum behavior \cite{jak}.
The trace anomaly is related to the divergent part of effective
action, in the absence of boundaries in odd spacetime dimensions
the conformal anomaly is absent \cite{Birrell} (see also the
appendix of the present paper). Under the conformal transformation
$g'_{ik}=\Omega ^{2}g^{{\mathrm{R}}} _{ik}$, the $\varphi
_{{\mathrm{R}}}$ field will change by the rule
\begin{equation}
\varphi (x')=\Omega ^{-3/2}\varphi _{{\mathrm{R}}}(x'),
\label{phicontr}
\end{equation}
where the conformal factor is given by expression (\ref{Omega}).
The scalar field $\varphi _{{\mathrm{R}}}(x')$, satisfy following
mixed boundary condition
\begin{equation}
(A_R+B_R \acute{n}_{R}^{l}\nabla'_{l})\varphi _{{\mathrm{R}}}=0,
\hspace{5mm}\xi=a, \hspace{5mm}\acute{n}_{R}^{l}\nabla'_{l}
=\delta^{l}_{1} \label{bounconr}
\end{equation}
 Now by comparing boundary conditions (\ref{boundcond}) and
(\ref{bounconr})   and taking into account Eq. (\ref{phicontr}),
one obtains the relation between the coefficients in the boundary
conditions:
\begin{equation}\label{relcoef}
A=\frac{1}{\Omega }\left(
A_{{\mathrm{R}}}+\frac{3}{2}B_{{\mathrm{R}}} n^l\nabla _l \Omega
\right) , \quad B=B_{{\mathrm{R}}}, \quad x\in S.
\end{equation}
To evaluate the expression $n^l\nabla _l \Omega $ we need the
components of the normal to $S$ in coordinates $x^i$. They can be
found by transforming the components $n'^l=\delta ^{l}_1/\Omega $ in
coordinates $x'^{i}$:
\begin{equation}\label{normal}
n^l=\left( 0,\frac{a}{\alpha }(\cos \theta -r/\alpha ), -\frac{a
}{\alpha r }\sin \theta ,0,0 \right) .
\end{equation}
Now it can be easily seen that $n^l\nabla _l \Omega =-\sqrt{\alpha
^2-r^2}/\alpha ^2$ and, hence, the relation between the Robin
coefficients in the Rindler and dS problems takes the form
\begin{equation}\label{relcoef1}
A=\frac{aA_{{\mathrm{R}}}}{\sqrt{\alpha
^2-r^2}}-\frac{3}{2}\frac{aB_{{\mathrm{R}}}}{\alpha ^2}, \quad
B=B_{{\mathrm{R}}} .
\end{equation}
 As for the energy-momentum tensor the spatial part is anisotropic, the
corresponding part in coordinates $x^i$ is more complicated:
\begin{eqnarray}
\left\langle T_{i}^{k}\left[ g_{lm},\varphi \right]\right\rangle
 &=& \Omega ^{-5}\left\langle T_{i}^{k}\left[
g^{{\mathrm{R}}}_{lm},\varphi _{{\mathrm{R}}}\right]\right \rangle
, \quad i,k=0,3,4, \label{tikv1}\\
\left\langle T_{1}^{1}\left[ g_{lm},\varphi \right]\right\rangle
 &=&\frac{(\cos \theta -r/\alpha )^2}{\Omega ^{7}}\left\langle
T_{1}^{1}\left[ g^{{\mathrm{R}}}_{lm},\varphi
_{{\mathrm{R}}}\right]\right\rangle  \nonumber \\
&& +\frac{1-r^2/\alpha ^2}{\Omega ^{7}} \sin ^2\theta \left\langle
T_{2}^{2}\left[ g^{\mathrm{R}}_{lm},\varphi
_{{\mathrm{R}}}\right]\right\rangle , \label{tikv2}\\
\left\langle T_{1}^{2}\left[ g_{lm},\varphi \right]\right\rangle
 &=&\frac{(r/\alpha -\cos \theta )\sin \theta }{r\Omega
^{7}}\left\{ \left\langle T_{1}^{1}\left[
g^{{\mathrm{R}}}_{lm},\varphi _{{\mathrm{R}}}\right]\right\rangle
-\left\langle T_{2}^{2}\left[ g^{{\mathrm{R}}}_{lm},\varphi
_{{\mathrm{R}}}\right]\right\rangle \right\} ,\label{tikv3}\\
\left\langle T_{2}^{2}\left[ g_{lm},\varphi \right]\right\rangle
 &=&  \frac{ 1-r^2/\alpha ^2}{\Omega ^{7}} \sin ^2\theta
\left\langle T_{1}^{1}\left[ g^{{\mathrm{R}}}_{lm},\varphi
_{{\mathrm{R}}}\right]\right\rangle \nonumber \\
&& + \frac{(r/\alpha -\cos \theta  )^2}{\Omega ^{7}} \left\langle
T_{2}^{2}\left[ g^{{\mathrm{R}}}_{lm},\varphi
_{{\mathrm{R}}}\right]\right\rangle  ,\label{tikv4}
\end{eqnarray}

\section{Conclusion} \label{secconc}

In the present paper we have investigated the surface Casimir
densities in dS spacetime for a conformally coupled massless
scalar field which satisfies the Robin boundary condition
(\ref{boundcond}) on a hypersurface described
 by equation (\ref{sur}). The coefficients in the boundary condition are
  given by relations (\ref{relcoef1}) with constants $A_{{\mathrm{R}}}$ and
   $B_{{\mathrm{R}}}$ and, in general, depend on the point of the hypersurface.
  The latter is the conformal image of the flat boundary moving by uniform proper
  acceleration in the Minkowski spacetime. We have assumed that the field in
  dS spacetime is in the state conformally related to the Fulling-Rindler vacuum.
  The energy-momentum tensor in  dS spacetime is generated from the corresponding
   results in the Rindler spacetime by using the standard formula for the
    energy-momentum tensors in conformally related problems in combination with
  the appropriate coordinate transformation. The Rindler energy-momentum tensor
     is taken from Ref. \cite{set5}, where the general case of the curvature
       coupling parameter is considered.  For a minimally coupled scalar field, the surface
energy-momentum tensor induced by quantum vacuum effects
corresponds to a source of a cosmological constant type located
on the plate and with the cosmological constant determined by the
proper acceleration of the plate. By using the conformal relation
between the Rindler and dS spacetimes and the results from
\cite{Saha02}, in Ref. \cite{set} the vacuum expectation value of
the bulk energy-momentum tensor is evaluated for a conformally
coupled scalar field which satisfies the Robin boundary condition
on a curved brane in dS spacetime.  By making use the same
technique and the conformal properties of the surface
energy-momentum tensor, from the results of the \cite{setsahr} we
have obtained the surface vacuum energy-momentum tensor induced on
the brane in dS spacetime, which is a conformal image of a
uniformly accelerated plate in the Minkowski spacetime. As it has
been shown recently in \cite{Saha04d}(see also \cite{set}), the
surface densities induced by quantum fluctuations of bulk fields
can serve as a natural mechanism for the generation of
cosmological constant in braneworld models of the Randall-Sundrum
type with the value in good agreement with recent cosmological
observations.
\section{Appendix}
As we have seen in previous sections the vacuum expectation values
of the surface energy-momentum tensor contain pole and finite
contributions. The remaining pole term is a characteristic feature
for the zeta function regularization method and has been found for
many other case of boundary geometries. In the conformally coupled
case, fluctuations of the stress tensor trace is as
\begin{equation}
<T^{i}_{i}(x)>=cK(x), \label{trac}
\end{equation}
where $c$ is a constant, and
\begin{equation}
K(x)=\zeta(s|A)(x)|_{s=0},\label{zet}
\end{equation}
here $\zeta(s|A)$ is the zeta function related to an elliptic
operator $A$ \cite{zerbi}. One can represent the zeta function as
following
\begin{equation}
\zeta(s|A)=\frac{1}{\Gamma (s)}\int_{0}^{\infty} dt t^{s-1}
K(t),,\label{zet1}
\end{equation}
where
\begin{equation}
K(t)=( \frac{1}{4\pi
t})^{3/2}\sum_{k}\exp(-\lambda_{k}t)\label{heat},
\end{equation}
is the heat kernel in four dimension, the $\lambda_{k}$'s are the
one-particle energies with the quantum number $k$. Now the
ultraviolet divergencies of the vacuum energy are determined from
the behaviour of the integrand in Eq.(\ref{trac}) at the lower
integration limit and, hence, from the asymtotic expansion of the
heat kernel for $t\rightarrow o$
\begin{equation}
K(t)\sim ( \frac{1}{4\pi t})
^{3/2}\sum_{k=0,1/2,1,...}B_{k}t^{k}.\label{asym}
\end{equation}
This expansion is known for a very general manifold, if the
underling manifold is without boundary, only coefficients with
integer numbers enter, otherwise half integer power of $t$ are
present. The $B_k$ are given by
\begin{equation}
B_{k}=\int_{M}dv a_{k}(x)+\int_{\partial M}ds
c_{k}(y),\label{coef}
\end{equation}
the Seely-de Witt coefficients $a_{k}$ vanish for half-odd
integers, these coefficients are independent of the applied
boundary condition, but the coefficients do depend on the spin of
the field in equation \cite{{Birrell},{blu},{setal}}. The
coefficients $c_{k}$ are functions of the second fundamental form
of the boundary (extrinsic curvature), the induced geometry on
the boundary (intrinsic curvature), and the nature of boundary
condition imposed. The simplest first of $a_{k}$ and $c_{k}$
coefficients for a manifold with boundary are given in
\cite{Birrell}.

\end{document}